\documentclass[11pt]{article}

\usepackage{fullpage}
\usepackage{amsmath}
\usepackage{amssymb}
\usepackage{mathrsfs}
\usepackage{setspace}
\usepackage{bbm}
\usepackage{dsfont}

\usepackage{graphics}

\usepackage[latin1]{inputenc}
\usepackage[pdftex]{graphicx}

\usepackage{color}
\usepackage[american]{babel}

\newcommand{\be}{\begin{equation}}
\newcommand{\ee}{\end{equation}}
\newcommand{\bes}{\begin{equation*}}
\newcommand{\ees}{\end{equation*}}
\newcommand{\bea}{\begin{eqnarray}}
\newcommand{\eea}{\end{eqnarray}}
\newcommand{\beas}{\begin{eqnarray*}}
\newcommand{\eeas}{\end{eqnarray*}}

\newcommand{\pr}[1]{\left(#1\right)}

\begin{document}
\numberwithin{equation}{section}
{
\begin{titlepage}
\begin{center}

\hfill \\
\hfill \\
\vskip 0.75in



{\Large \bf On the Universality of Inner Black Hole Mechanics and Higher Curvature Gravity}\\

\vskip 0.4in

 { Alejandra Castro${}^a$, Nima Dehmami${}^b$, Gaston Giribet${}^c$ and David Kastor${}^d$}\\

\vskip 0.3in

{\it$^a$ Center for the Fundamental Laws of Nature, Harvard University, Cambridge 02138, MA USA }  \vskip .5mm              
{\it $^b$ Physics Department, Boston University, Boston 02215, MA USA }\vskip .5mm  
{\it $^c$ Physics Department, University of Buenos Aires and CONICET, C1428ZAA, Argentina}\vskip .5mm  
{\it $^d$ Department of Physics, University of Massachusetts, Amherst, MA 01002, USA }

\vskip 0.3in

\end{center}

\vskip 0.35in

\begin{abstract} 
Black holes are famous for their universal behavior. New thermodynamic relations have been found recently for the product of gravitational entropies over all the horizons of a given stationary black hole. This product has been found to be independent of the mass for all such solutions of Einstein-Maxwell theory in $d=4,5$.
We study the universality  of this mass independence by introducing a number of possible higher curvature corrections to the gravitational action.  We consider finite temperature black holes with both asymptotically flat and (A)dS boundary conditions. Although we find examples for which mass independence of the horizon entropy product continues to hold, we show that the universality of this property fails in general.   We also derive further thermodynamic properties of inner horizons, such as the first law and Smarr relation, in the higher curvature theories under consideration, as well as a set of relations between thermodynamic potentials on the inner and outer horizons that follow from the horizon entropy product, whether or not it is mass independent.

\end{abstract}

\vfill

\noindent \today

\end{titlepage}
}

\newpage
\tableofcontents

\section{Introduction}

A fascinating feature of  black hole thermodynamics is its apparent universal scope beyond general relativity.  While the status of the second law in higher curvature theories remains unresolved, the geometric expression for black hole entropy and its role in the first law in these theories are well established  \cite{Wald:1993nt,Iyer:1994ys}.
Moreover, in certain examples accessible via string theory a precision accounting for the black hole entropy including higher curvature corrections can be made in terms of microscopic degrees of freedom  (see {\it e.g.} \cite{Sen:2007qy} for a review).
Although our understanding of its statistical underpinnings remains incomplete, it seems clear that the universal laws of black hole thermodynamics  reflect important features of the underlying quantum mechanical degrees of freedom. 
In classical general relativity, it has been observed that certain additional thermodynamic relations also appear to be universal \cite{Cvetic:2010mn,Castro:2012av,Detournay:2012ug} and may provide further insight into the quantum physics of black holes. Here we will test the universality of  these new  thermodynamic relations by investigating to what extent they do, or do not, hold in higher curvature gravity theories of gravity.

These new relations are novel in the sense that they involve thermodynamic quantities defined at different horizons, {\it e.g.} the inner and outer horizons of rotating and/or charged black holes.  For example, it has been observed that for all known stationary solutions of Einstein-Maxwell theory in $d=4,5$
\be\label{eq:areas}
\prod_i A_i = {\cal Q}~,
\ee
where the $A_i$ are the areas of {\it all} the Killing horizons
and the quantity  ${\cal Q}$ depends on conserved charges such as the charge and angular momentum, but is remarkably independent of the black hole mass.
This implies, for example, that a non-extreme black hole will have the same value for the horizon area product as the extreme black hole with the same charge and angular momentum.
Since we are dealing with solutions to Einstein gravity at this point, black hole entropy is simply proportional to the horizon area and 
(\ref{eq:areas}) may be equivalently written as
\be\label{eq:main}
\prod_i (4G S_i) = {\cal Q} ~,
\ee
and interpreted as relating thermodynamic quantities defined at the different horizons.  It turns out that these relations are highly robust in Einstein gravity,  
having been tested for a wide class of solutions as we will review in  section \ref{sec:PA} below. 
Equation \eqref{eq:main} has important implications for the statistical interpretation of  black hole entropy.  In the asymptotically flat case, equation \eqref{eq:main} may be used in conjunction with the thermodynamic relations for the inner horizon to show that  the Bekenstein-Hawking entropy can always be written in terms of a  Cardy-type  formula \cite{Chen:2012mh,Chen:2013rb,Castro:2013kea}.   This provides additional evidence for a CFT description of black hole microstates \cite{Strominger:1997eq,Larsen:1997ge,Guica:2008mu,CMS}.   However, so far no geometric proof  has been found for equation \eqref{eq:areas}, which makes the ultimate significance of these observations uncertain. 

In the absence of a geometric understanding, it makes sense to explore phenomenologically whether a horizon area or entropy product relation holds universally in arbitrary gravity theories. 
Towards this goal, we will consider a number of different higher curvature modifications of the Einstein-Hilbert action. With such modifications, of course, the entropy will not generally be proportional to the horizon area, and the product relations for the horizon areas and entropies will no longer be equivalent.   
Since black hole entropy can be viewed as a Noether charge \cite{Wald:1993nt}, it might be more natural to expect that the entropy product formula \eqref{eq:main}, rather than the product of the areas  \eqref{eq:areas}, would be the correct generalization for modifications to Einstein gravity\footnote{We will return to this point in section \ref{sec:fr}. The argument is that if we view the additional higher curvature terms as arising from an infinite series of higher derivative corrections, then the area of the black hole is not well defined due to ambiguities in field redefinitions. }. 
However, our results do not bear out this intuition.  Although we will find examples in which an entropy product formula holds, which we will consider ``successes", these are rather non-generic situations in which the entropy remains proportional to the horizon area.  It is straightforward to find ``failures", in which the product of the horizon entropy depends on the mass.  In all the examples we consider, on the other hand, we find that the product of the horizon areas remains independent of the black hole mass.

Constructing exact black hole solutions in higher derivative theories is itself a difficult task, and this makes testing the validity of product formulas for the black hole horizon entropy or area challenging.  In the following we will make some simplifications to alleviate our task, and they fall into three broad classes\footnote{Throughout the text we will make a distinction between ``higher derivative'' and ``higher curvature'' corrections. Higher derivative theories will denote modifications to the Einstein-Hilbert action that make the equations of motion of higher order, {\it i.e.} involving third, or higher, derivatives of the metric.   Higher curvature theories are modifications to Einstein theory such that  the equations of motion are not necessarily of higher order.}

\begin{enumerate}
\item {\it Einstein manifolds}:  The simplest scenario is to consider corrections to the Einstein-Hilbert action such that the additional higher curvature terms either do not affect  Einstein's equations or the equations of motions still admit Einstein manifolds as  solutions. This will be the case for $d=4$, where the higher curvature terms considered are the Gauss-Bonnet invariant and the Weyl  tensor squared. 
\item {\it Lovelock theories}: Simplifications also occur if we include higher curvature terms such that the   equations of motion remain of second order. This is the defining feature of Lovelock theories \cite{Lovelock:1971yv}. This simplification allows for the construction of static black holes with the desired features to test the entropy and area product relations.
\item {\it $f(R)$ gravity}:  The higher derivative corrections involve only functions of the Ricci scalar. Even though the equations of motion are of higher order, for those solutions that have constant Ricci curvature, the equations can be solved in closed form. This resembles very closely the effects of higher curvature corrections in AdS$_3$ gravity, and hence our results mimic those of the BTZ black hole. 
\end{enumerate}
One feature that is absent for each item in this list is a consistent treatment of higher derivative corrections for the matter fields, in addition to the gravitational sector.   Our explicit examples of charged black holes will include only the Maxwell term in the action; there are no couplings between higher curvature terms and matter fields.  Hence our results do not connect with string theory construction of BPS solutions in higher derivative theories.

  The organization of the paper is as follows. In section \ref{sec:PA} we will review what is known about entropy product formulas for black holes in Einstein-Maxwell theory. In section \ref{sec:uls}  and \ref{sec:luf} we study certain thermodynamic properties of black holes in higher curvature gravity. Besides evaluating the horizon entropy and area products, we will also discuss the thermodynamics of the Cauchy horizons, including generalizations of the Smarr relations.  The evident difference between our ``successful" examples in section \ref{sec:uls} and the ``failures" in section \ref{sec:luf} is the scaling of  the Wald entropy with the area of the black hole. For our successes it is systematically the case that the entropy is proportional to the area, where the higher derivative corrections are merely encoded in renormalizing Newton's constant. This is not a generic property of Wald entropy formula, and our failures illustrate this explictly. 
In section \ref{sec:disc} we will discuss further implications of the lack of universality we have found for the entropy product relation \eqref{eq:main} and also mention future directions for investigation. In appendix \ref{sec:defn} we collect definitions which are used throughout the paper. In appendix \ref{app:A} we generalize the discussion in section \ref{sec:luf} to higher dimensions and (A)dS solutions in Lovelock gravity. Finally, in appendix \ref{app:YM} we discuss the six dimensional Yang monopole solution as another bizarre occurrence of \eqref{eq:main}.
     
\section{Products of Areas in Einstein-Maxwell Gravity}\label{sec:PA}

It is remarkable that all known solutions in $d=4,5$ Einstein-Maxwell gravity satisfy the product relation \eqref{eq:main}; this includes solutions that are either asymptotically flat, AdS or dS. There are as well some  similar observations in $d=3$ --which we will review in section \ref{sec:BTZ}-- and  a few generalizations to $d\geq 6$.\footnote{Classical solutions in $d\geq6$ are less well explored and hence a complete classification is lacking. It is known that ``no-hair'' theorems are not valid, and therefore we do not expect universal properties for these theories.}  For the purposes of understanding the more sophisticated cases in higher derivative gravity, it is instructive to review which ingredients  go into these universal properties.

 \subsection{Asymptotically Flat Solutions}\label{sec:AF}
 
For a large class of black holes, it has been observed that the product $A_+ A_-$ is independent of the mass of the black hole \cite{Curir1,Curir2,Larsen:1997ge,Ansorg:2008bv, Ansorg:2009yi,Meessen:2012su,Castro:2012av,Visser:2012zi}. Here $A_\pm$ are the areas inner and outer horizons.  While a robust property, we emphasize that there is no derivation that explains why the product is protected.

An alternative phrasing of this observation is
\be\label{eq:aa}
S_+S_- = (S_{\rm ext})^2 ~,
\ee
where $S_{\pm}=A_{\pm} /(4G)$, and where $S_{\rm ext}$ is the entropy of the extremal ($T=0$) black hole. The right hand side of this expression is independent of any modulus, which hints at the fact the product might be the result of a fixed point of an appropriate ODE.   Further, it has been shown that for those solutions supported by neutral scalar fields, the scalars satisfy a similar geometric mean property \cite{Meessen:2012su}. This has an interesting resemblance with the attractor mechanism, which would be interesting to sharpen, if possible.

To be clear about the scope of the relation (\ref{eq:aa}), let us review what are the precise properties shared by the solutions:\footnote{It is assumed as well that the solution is stationary. For a discussion on time dependent backgrounds see \cite{Faraoni:2012je}.}
 \begin{enumerate}
 \item These are solutions to Einstein-Maxwell theory, and its supergravity generalizations. Further, there is no classical hair in these theories.  
 \item  The black hole is at finite temperature (non-extremal) and has a smooth extremal limit where $S_+=S_-$. Notice that, for instance, Schwarzschild solution does not fall into this category. 
 \item The only two Killing horizons are the inner and outer horizon. This is the case for all solutions considered in $d=4,5$. For solutions in $d>5$, equation \eqref{eq:aa} has to be modified to accommodate additional Killing horizons.
 \item The black hole is single centered.\footnote{There might be interesting generalizations to multi-centered black holes. The difficulty of course is to find multi centered configurations that satisfy the above requirements.} 
 \item The topology of the horizon for known solutions is irrelevant.  In particular, the relation \eqref{eq:aa} applies to black rings and strings \cite{Meessen:2012su,Castro:2012av} in addition to spherical black holes. Whether this is a universal feature or not is  also something that can be investigated by considering higher curvature black holes.
 \end{enumerate}
In the following sections we will consider examples which satisfy the above properties with the obvious exception that they will not be solutions to Einstein-Maxwell theory. 

 \subsection{Asymptotically (A)dS Solutions}

For either asymptotically AdS or dS black holes, \eqref{eq:aa} has to be modified. The observation made in  \cite{Cvetic:2010mn} is that  the product of the areas of {\it all} possible poles of the radial metric function is independent of the mass of the black holes. These poles must generate a Killing horizon with finite area; however the roots are not necessarily physical since the prescription given in \cite{Cvetic:2010mn} requires including virtual horizons where the area can be a complex number. 

Following \eqref{eq:aa} we will rewrite \eqref{eq:main} as
\be\label{eq:psads}
\prod_{i=1}^{n}S_i = \prod_{i=1}^n S_{(i,\rm ext)}~,
\ee
where $S_{(i,\rm ext)}$ is the entropy of each virtual horizon at zero Hawking temperature.\footnote{Actually any extremal limit that the solution admits---where two poles coincide and hence the surface gravity of the resulting horizon vanishes---will be consistent with \eqref{eq:psads}.} We are merely emphasizing in \eqref{eq:aa} and \eqref{eq:psads} that having the product of entropies being independent of the mass is equivalent to the statement of it being independent of the temperature of the black hole. 
 This product relation has been tested for a wide class of solutions \cite{Toldo:2012ec, Klemm:2012vm,Visser:2012wu,Gnecchi:2012kb}. All of these configurations  share the following properties:
 \begin{enumerate}
 \item These are solutions to Einstein-Maxwell gravity including a cosmological constant, and its gauged supergravity generalizations. 
  \item The product involves all possible bifurcate horizons, i.e. all roots of the appropriate radial direction.  Generically, this involves complex roots, and hence unphysical horizons.  
 \item  The black hole has a smooth extremal limit when two (or more) roots coincide. 
 \item The black hole is single centered.
 \end{enumerate}
(A)dS-Schwarzschild does fall into this category in the sense that the geometric mean of entropies satisfies \eqref{eq:psads}.  However, for (A)dS-Schwarzschild the extremal entropy $S_{\rm ext}$---which is the Narai limit---depends on the mass of the black. This resembles closely the case discussed in appendix \ref{app:YM} for the six dimensional Yang monopole.

\section{Universal but Limited ``Success"}\label{sec:uls}

In this section, we will begin our discussion of horizon entropy products in higher derivative theories on a positive note by focussing on examples that do work, {\it i.e.} in which the horizon product is independent of the mass as in Einstein gravity.  We will see that such examples have a universal character, in that they arise in a wide class of theories.  However, this universality is limited by a key non-generic feature shared by all the examples.  Namely,  that all higher derivative corrections to the entropy are proportional to the horizon area.

\subsection{BTZ  in Higher Derivative Gravity }\label{sec:BTZ}

BTZ black holes  \cite{Banados:1992wn,Banados:1992gq} considered as solutions to higher derivative theories in three dimensions provide important examples.  
In Einstein gravity,  a general argument based on representations of the asymptotic symmetry group \cite{Brown:1986nw,Strominger:1997eq} (not depending on supersymmetry or string theory)  provides an accounting of the BTZ black hole entropy in terms of quantum mechanical microstates.  The mass independence of the horizon entropy product for BTZ black holes consequently has a clear interpretation in terms of quantized degrees of freedom.

BTZ black holes are locally identical to AdS spacetimes.  It follows that any higher derivative theory having a constant negative curvature vacuum also has BTZ black hole solutions.  Equality of the geometrical and quantum statistical entropies for BTZ black holes in higher derivative theories was established in \cite{Saida:1999ec}.  
 Further, the geometrical entropy remains proportional to the one-dimensional area (length) of the horizon cross-section, but rescaled by a factor that depends on the couplings of the higher derivative terms in the gravitational action \cite{Kraus:2005vz,Gupta:2007th}. The argument is rather simple and it goes as follows. The Wald entropy is given by
 \be
S_+= 2\pi \int_{\Sigma^{+}} d\phi \sqrt{g_{\phi\phi}}  {\partial {\cal L}_3\over \partial R_{\mu\nu\lambda\rho}}\epsilon_{\mu\nu}\epsilon_{\lambda\rho} ~.
 \ee
 The Riemann tensor can always be expressed in terms of the Ricci tensor in  $d=3$, which allows us to write ${\cal L}_3={\cal L}_3(g_{\mu\nu},R_{\mu\nu})$. This simplifies the Wald entropy to
 \be
 S_+= 2\pi \int_{\Sigma^{+}} d\phi \sqrt{g_{\phi\phi}} {\partial {\cal L}_3\over \partial R_{\mu\nu}}g^{\alpha\beta}\epsilon_{\mu\beta}\epsilon_{\nu\alpha}~.
 \ee
 Since BTZ is locally equivalent to $AdS_3$,  the local variations in the integrand are insensitive to the black hole, and hence we can write
\be
S_+= {A_+\over 4G_{\rm eff}} ~,\quad {1\over G_{\rm eff}}= {16\pi \over 3} g^{\mu\nu}{\partial {\cal L}_3\over \partial R_{\mu\nu}}~.
\ee
The same argument holds at the inner horizon. We emphasize that $G_{\rm eff}$ is a function of the AdS$_3$ radius and the couplings of the theory, but insensitive to the black hole parameters. Therefore, if there is no diffeomorphism anomaly, the horizon entropy product remains independent of the mass  \cite{Detournay:2012ug}.  BTZ solutions of such higher derivative theories thus provide a wide class of examples in which equation (\ref{eq:main}) holds.

It is worth noting, however, that higher derivative theories in three dimensions may possibly have additional, non-BTZ black hole solutions as well.  It has not been established whether, or not, equation (\ref{eq:main}) will hold for all black holes in these theories.   In the specific case of New Massive Gravity (NMG) \cite{Bergshoeff:2009hq,Clement:2009gq}, which includes certain curvature squared terms in the action, equation (\ref{eq:main}) has been shown to hold for warped $AdS_3$ black holes in NMG  as well \cite{Detournay:2012ug}.

\subsection{Charged Black Holes in $f(R)$ Gravity}

One might imagine that the combined simplifications of three dimensional geometry and the constant curvature nature of BTZ black holes, would make the broad class of examples of the last section unique. However, a similar degree of simplification arises for certain solutions of so-called $f(R)$ theories in which the gravitational Lagrangian depends only on the scalar curvature.  Our discussion here follows \cite{Sheykhi:2012zz} (see also \cite{Sotiriou:2008rp,Moon:2011hq}).  Explicitly, let us  consider an action of the form
\be\label{eq:acfr}
S={1\over 16\pi G} \int d^4x \sqrt{-g}\, \left(R+f(R)-F_{\mu\nu}F^{\mu\nu}\right)~,
\ee
where $f(R)$ is an arbitrary function of the scalar curvature, and we introduce the notation $f'(R)=\partial f(R)/\partial R$. 
These theories can be alternatively written as 
\be\label{eq:acfrp}
S={1\over 16\pi G} \int d^4x \sqrt{-g}\, \left(\Phi R+V(\Phi)-F_{\mu\nu}F^{\mu\nu}\right)~,
\ee
where we introduced an auxiliary field $\Phi$. The equations of motion derived from \eqref{eq:acfrp} are equivalent to those in \eqref{eq:acfr}, were one identifies $\Phi=1+ f'(R)$. Further, integrating out $\Phi$ reproduces again \eqref{eq:acfr}. Several of the properties we will encounter below are simple due to the fact that $f(R)$ gravity is equivalent  to Einstein gravity coupled to matter.

A simple class of static, spherically symmetric solutions can be obtained if the theory admits a constant curvature vacuum \cite{Moon:2011hq}.  If we denote the constant  scalar curvature by $R=R_0$ then the solution is given  by
\be
ds^2=-N(r)dt^2+{dr^2\over N(r)} +r^2d\Omega_2^2~,\quad F_{tr}={q\over r^2}~,
\ee
with the metric function
\be\label{metricfunction}
N(r)=1-{2\mu\over r}+{q^2\over r^2} {1\over (1+f'(R_0))}-{R_0\over 12}r^2=-{R_0\over 12 r^2}\prod_{i=1}^{4}(r-r_i)~.
\ee
These solutions are clearly very similar to RN(A)dS spacetimes  and represent, for a given value of $R_0$,  asymptotically (A)dS black holes over a range of the $(\mu,q)$ parameter space.  We will refer to the four roots $r_i$ of the equation $N(r)=0$ as horizons, independent of whether, or not, they represent physical black hole or Cauchy horizons.
There are generalizations of these solutions, involving modifications to the matter action as well, in dimensions $d=4k$ where $k=1,2,\dots$ \cite{Sheykhi:2012zz}. However, for our purposes it is sufficient to focus on $d=4$.  Our results  extend straightforwardly to the higher dimensional cases.

The mass and electric charge of these constant curvature solutions to $f(R)$ gravity  are given by
\bea
M={\mu\over G} (1+f'(R_0))~,\quad Q={1\over \sqrt{2G}}{q\over \sqrt{1+f'(R_0)}}~,
\eea
where the normalizations have been fixed such that in the limiting case of vanishing scalar curvature, $f'(R_0)=0$, the conventions used in section \ref{sec:4d} are recovered. Following the definitions \eqref{genpot} and \eqref{elecpot}, the temperature and chemical potential of each horizon $r_i$ (either real or complex) is given by
\bea\label{eq:T1}
T_i={1\over 4\pi r_i}\left(1-{q^2\over r^2_i} {1\over (1+f'(R_0))}-{R_0\over 4}r^2_i\right)~,\quad \Phi_i=\sqrt{2\over G}{q\over r_i} \sqrt{ 1+f'(R_0))}~,
\eea
while the Wald entropy \cite{Wald:1993nt} is given by
\be\label{eq:T2}
S_i
={A_i\over 4G}(1+f'(R_0)).
\ee
The fact that, like the BTZ black holes considered in section (\ref{sec:BTZ}), the entropy for these black holes including higher derivative corrections is proportional to the horizon area is a key feature.

Given the expressions above,  it is straightforward to check that the first law and Smarr relation
\bea\label{eq:T3}
dM&=&T_idS_i+\Phi_i dQ~,\\    
\label{eq:T4}
M &=& 2T_iS_i +\Phi_i Q +{R_0\over 12 G}(1+f'(R_0))r_i^3~,
\eea
are satisfied for each horizon.  We speculate that in the asymptotically AdS case, the final term in (\ref{eq:T4}) should be the Casimir energy of the boundary CFT.  It would also be of interest to see whether this term follows via an overall scaling from an extended first law including variations in the gravitational couplings,  as do analogous terms in the Smarr formulas for Einstein gravity with $\Lambda\neq 0$ \cite{Kastor:2009wy} and Lovelock gravity \cite{Kastor:2010gq}, and whether this term may be interpreted as the thermodynamic volume of the black hole as in \cite{Cvetic:2010jb,Dolan:2013ft}.

It follows directly from equation (\ref{metricfunction}) that the product of horizon areas for charged $f(R)$ black holes is given by $\prod_{i=1}^4 A_i
= \left({24 G\over R_0}\right)^2(4\pi Q)^4$
which is clearly independent of the mass. Similarly, because the entropy in (\ref{eq:T2}) is proportional to the horizon area, the product of the entropies is also independent of the mass,
\bea
\prod_{i=1}^4 S_i= \left({24\over G R_0}\right)^2 \left ((1+f'(R_0))\pi Q\right)^4.
\eea
If we consider the limit of $R_0=0$ while keeping $f'(R_0)\neq0$, then the solution is asymptotically flat  and there are only two roots ($r_\pm$). The product of the entropies gives
\be
S_+S_-=4\pi^2 Q^4 (1+f'(R_0))^2 = (S_{\rm ext})^2~,
\ee
where $S_{\rm ext}$ is the entropy of the extremal black hole. 

The class of charged $f(R)$ black holes with constant scalar curvature satisfies the product relation \eqref{eq:main}. The origin of this success is clear. The entropy \eqref{eq:T2} is corrected only by a multiplicative factor that is independent of the horizon geometry and could, in fact, be absorbed into a correction to Newton's constant 
\be
G~\to~ G_{\rm eff}\equiv {G\over 1+f'(R_0)}~.
\ee
As in the three dimensional case above, it should be noted that there may also be additional black hole solutions to $f(R)$ gravity ({\it e.g.} without constant scalar curvature) that do not satisfy  \eqref{eq:main}.
The situation in $f(R)$ gravity will be contrasted below with the examples of charged black holes in Gauss-Bonnet and more general Lovelock gravity theories, for which the higher curvature corrections to the entropy cannot be absorbed in a renormalization of the Newton's constant.

\section{Limited but Universal ``Failure"}\label{sec:luf}

In this section we will construct, in very simple setups, counterexamples to the entropy product relation \eqref{eq:main}. For simplicity we will focus only on Lovelock theories in $d=4,5$ and the corresponding asymptotically flat black holes; generalizations to higher dimensions and inclusion of a cosmological constant can be found in appendix \ref{app:A}. 

With respect to the assumptions listed in section \ref{sec:AF}, the only modification is to the gravitational Lagrangian. In particular, the analytic properties of the  black holes are unchanged, and the extremal limits are well defined. Still, we will see explicitly that the universal features of Einstein-Hilbert action breakdown quickly after minor modifications of the action.

\subsection{Two Four Dimensional Counterexamples}\label{sec:4d}

We begin by considering two simple examples of higher derivative theories in $d=4$, which share the feature that solutions to the vacuum Einstein equations are solutions to the full theory.   Take the action to have the general form
\be\label{eq:ba}
S={1\over 16\pi G} \int d^4x \sqrt{-g}\, \left({\cal L}_1+{\cal L}_2\right)+  \int d^4x \sqrt{-g}\, {\cal L}_{\rm matter} ~,
\ee
where ${\cal L}_1=R$ and ${\cal L}_2$ encodes the higher derivative gravitational interactions.

 For our first example, we take the Gauss-Bonnet combination of quadratic curvature terms
 \be\label{eq:Max}
{\cal L}_{2}={\cal L}_{\rm GB}=\alpha \left(R_{\mu\nu\lambda\rho}R^{\mu\nu\lambda\rho}-4 R_{\mu\nu}R^{\mu\nu}+R^2\right) ~,
 \ee
 and include also the electromagnetic field with  ${\cal L}_{\rm matter}= -{1\over 4}F_{\mu\nu}F^{\mu\nu}$.  The integral of ${\cal L}_{\rm GB}$ in $d=4$ is the topologically invariant Euler character, and therefore the equations of motion in this theory are identical to those of Einstein-Maxwell theory.  The black hole entropy, however, will  include a correction term proportional to the Gauss-Bonnet coupling.

Einstein-Weyl gravity provides a second simple example, with
\be\label{eq:weyl}
{\cal L}_{2}= {\cal L}_{\rm W}=\alpha\, C_{\mu\nu\lambda\rho}C^{\mu\nu\lambda\rho}
 = \alpha\left(R_{\mu\nu\lambda\rho}R^{\mu\nu\lambda\rho}-2R_{\mu\nu}R^{\mu\nu}+{1\over 3}R^2\right)~,
\ee
and ${\cal L}_{\rm matter}=0$.  The contribution to the equations of motion from ${\cal L}_{\rm W}$ is proportional to the Bach tensor, 
$B_\mu{}^\nu= (\nabla_\rho\nabla^\sigma +{1\over 2}R_\rho{}^\sigma)C_{\mu\sigma}{}^{\nu\rho}$,
which can be shown to vanish for any Einstein space, {\it i.e.} one for which the Ricci tensor is proportional to the metric.
It follows that solutions to the vacuum Einstein equations form a subset of the solutions to Einstein-Weyl gravity.   In particular, the Kerr geometry is a solution to Einstein-Weyl gravity, while the full Kerr-Newman spacetime solves the equations of motion in our first example, which includes the Maxwell term as well as the Gauss-Bonnet term in the action.  For keeping track of dimensions, it is helpful to note that for both ${\cal L}_{\rm GB}$ and ${\cal L}_{\rm W}$ the coupling constant $\alpha$ has dimensions of $({\rm length})^2$.

\subsubsection{Kerr-Newman Black Hole}

We start by reviewing properties of the Kerr-Newman black hole.  The metric is given by
\be
ds^2={\Sigma\over {\Delta}}dr^2- {{\Delta}\over \Sigma}\left(dt -{a}\sin^2\theta\, d\phi\right)^2+ \Sigma d\theta^2+{ \sin^2\theta \over \Sigma} \left((r^2+{a}^2)\,d\phi-{{a}}\,dt\right)^2~,
\ee
where
\be
\Sigma= r^2+a^2\cos^2\theta~,\quad \Delta= r^2-2\mu r +a^2 +q^2~.
\ee
The Maxwell one-form is
\be
A= q {r\over \Sigma}(dt-a\sin^2\theta d\phi)~.
\ee
Here $F =dA$ and we taken the charge to be purely electric field. Straightforward generalizations exist with multiple $U(1)$ charges and with both electric and magnetic charges.

The presence of either choice for ${\cal L}_2$ does not modify the ADM charges,\footnote{Although this is the case for asymptotically flat solutions, conserved charges may be affected by the presence of ${\cal L}_2$ terms in asymptotically AdS spaces \cite{Kastor:2011qp}.} and hence the mass, angular momentum and electric charge are given by\footnote{Here 
$\Sigma_2=4\pi$ is the area of a 2-sphere, and our conventions for electric charge match with those used in \cite{Cvetic:2010mn,Castro:2012av}.}
\be\label{eq:qq}
M={\mu\over G} ~,\quad J=aM ~,\quad Q=\left({ \Sigma_2\over 8\pi G }\right)^{1/2} {q}~.
\ee
The locations of the inner and outer horizon $r_{\pm}$ are determined by the zeroes of $\Delta$. In terms of the parameters $(\mu,a,q)$ one finds that 
$r_++r_-=2\mu$ and $r_+r_- = a^2+q^2$.
Following the definitions in appendix \ref{sec:defn}, the horizon generating Killing vector are given by $\xi_\pm = \partial_t +\Omega_\pm \partial_\phi $,
where the angular velocity at each horizon are $\Omega_\pm=a/(r_\pm^2+a^2)$.
From \eqref{genpot} and  \eqref{elecpot}, the temperatures and electric potentials  of the inner and outer horizons are
\be\label{eq:cd}
T_\pm={r_+-r_-\over 4\pi (r_\pm^2 +a^2)}~,\quad \Phi_{\pm}=\left({ 8\pi \over\Sigma_2 G }\right)^{1/2} {q r_{\pm}\over r_\pm^2+a^2}~.
\ee

Even though ${\cal L}_2$ did not affect the value of the global charges or their conjugate potentials, the entropy of the black hole is sensitive to the presence of higher curvature terms in the gravitational action. With ${\cal L}_2={\cal L}_{\rm GB}$, the inner and outer  horizon entropies are given by
\be\label{eq:w4d}
S_\pm= {\pi\over G}\left(r_{\pm}^2+a^2+4\alpha\right)  ~.
\ee
The correction to the entropy from the Gauss-Bonnet term is proportional to the integral over a horizon cross-section of the scalar curvature of the induced metric on this surface \cite{Jacobson:1993xs}.  Since the horizon cross-section is a $2$-sphere, this integral is simply the Euler character of the sphere and is independent of the horizon radius.

The Kerr black hole ($Q=0$) is also a solution to Einstein-Weyl gravity.  Because $R_{\mu\nu}=0$ for Kerr, the $({\rm Weyl})^2$ interaction term (\ref{eq:weyl}) agrees with the Gauss-Bonnet term (\ref{eq:Max}) on-shell and hence leads to the same expression for the entropy (\ref{eq:w4d}). 
One then finds that the the entropy products in both theories are given explicitly in terms of the global charges by
\be\label{eq:prod4d}
  S_+S_-=4\pi^2 \left[J^2+ (Q^2-{2\alpha\over G})^2  + 4 M^2\alpha \right]   
\ee
which is clearly {\it not} independent of the mass. 

It is also instructive to examine the role played by the higher derivative terms  in the  thermodynamic properties of these solutions. The ordinary  first law of thermodynamics 
$dM=\pm T_{\pm} dS_{\pm} +\Omega_{\pm}dJ+\Phi_{\pm} dQ$ will still hold for both horizons.
However, we may also consider an extended first law, as proposed in \cite{Kastor:2010gq}, in which the higher derivative coupling $\alpha$ is varied as well
\be\label{eq:fthc}
dM= \pm T_{\pm} dS_{\pm} +\Omega_{\pm}dJ+\Phi_{\pm} dQ +\Theta_\pm d\alpha ~,
\ee
where the thermodynamic potential conjugate to $\alpha$ is defined to be $\Theta_\pm\equiv (\partial M/\partial\alpha){}_{S_\pm,J,Q}$.  
It follows from the form of the entropy (\ref{eq:w4d}) that for the Kerr(-Newman) solutions $\Theta_\pm$  is proportional to the horizon temperature,
\be
\Theta_\pm = \mp {4\pi\over G}T_\pm~.
\ee
Because $\alpha$ is dimensional, such an extension is necessary in order to obtain a Smarr formula from the first law via an overall rescaling of parameters  (see {\it e.g.} \cite{Kastor:2010gq}). It then follows that the Smarr formula for the Kerr(-Newman) spacetimes, considered as solutions to the higher derivative gravity theories with ${\cal L}_{\rm GB}$ or ${\cal L}_{\rm W}$ is given by
\be\label{eq:smarr4d}
M = 2(\pm T_{\pm} S_{\pm} +\Omega_{\pm}J)+\Phi_{\pm} Q + \Theta_{\pm} \alpha ~,
\ee
where the final term simply compensates for the additional term in the entropy  (\ref{eq:w4d}) relative to Einstein gravity.
It would be interesting to further understand the geometric significance of the potentials $\Theta_\pm$.

Despite the fact that entropy product \eqref{eq:prod4d} depends on the mass, it is worth emphasizing  that it is independent of the Smarr formula \eqref{eq:smarr4d}.  Therefore \eqref{eq:prod4d} implies a set of relations between the thermodynamic potentials at the two horizons. These are found by taking the differential of both sides of \eqref{eq:prod4d} and then using  the first law \eqref{eq:fthc} to eliminate $dS_\pm$;  requiring that the coefficients of $dM$, $dJ$, $dQ$ and $d\alpha$ are equal on both sides, which yields the relations
\bea\label{manyrelations}
{1\over T_+ S_+}-{1\over T_-S_-} &=& {32\pi^2 M\alpha\over S_+S_-}\cr
{\Omega_+\over T_+ S_+}-{\Omega_-\over T_-S_-} &=& -{8\pi^2 J\over S_+S_-}\cr
{\Phi_+\over T_+ S_+}-{\Phi_-\over T_-S_-} &=& -{16\pi^2 (Q^2-2\alpha/G)Q\over S_+S_-}\cr
{\Theta_+\over T_+ S_+}-{\Theta_-\over T_-S_-} &=& -{16\pi^2(M^2-Q^2+2\alpha/G)\over S_+S_-}
\eea
Some of these relations have been reported for Einstein-Maxwell black holes ({\it i.e.} with $\alpha=0$) in {\it e.g.} references \cite{Okamoto,Chen:2012mh}. 
The first relation, in particular, generalizes a well known relation between the two horizons in the $\alpha=0$ case when the entropy product is mass independent.
An understanding of the physical significance of the full set of  relations would require deeper intuition into the importance of  inner horizon thermodynamics.

Finally, our analysis did not include a cosmological constant ($\Lambda\neq 0$). Nevertheless, from the nature of the corrections here we speculate that the product relation \eqref{eq:main} for asymptotically (A)dS black hole will not be quantized in the presence of  higher derivative gravitational interactions ${\cal L}_{\rm GB}$ and/or ${\cal L}_W$.

\subsubsection{Comments on Field Redefinitions}\label{sec:fr}

We found that black holes solutions in Einstein-Maxwell-Gauss-Bonnet and Einstein-Weyl theories in 4D do not satisfy \eqref{eq:aa}. Curiously, however, the horizon areas do still satisfy the relation
\be\label{eq:areaprod}
A_+ A_- = (A_{\rm ext})^2~,
\ee
because neither the geometry, nor the conserved charges, are modified by the presence of the higher derivative terms.   It is unclear what physical significance should be attributed to this result.
One needs to be careful when discussing specific properties of the metric in higher curvature theories. As part of an effective field theory, ${\cal L}_2$ could be viewed as the first correction of an infinite series of higher curvature terms, and the coupling $\alpha$ controls the perturbative expansion.  In this context, the metric and matter field suffer ambiguities due to field redefinitions. In particular we could redefine the metric as
\bea\label{eq:redef}
g_{\mu\nu}\to \tilde g_{\mu\nu}= g_{\mu\nu}+a_1 \alpha R_{\mu\nu} + a_2 \alpha R g_{\mu\nu} + O(\alpha^2)~,
\eea
for arbitrary constants $a_{1,2}$. Such a transformation does not generally preserve the area of the black hole. However, one expects that the entropy should be invariant under \eqref{eq:redef} since it corresponds to a  Noether charge \cite{Jacobson:1993vj}.\footnote{This issue has been discussed in the content of computing the entropy of extremal black holes, including higher derivative corrections. The Wald entropy can then be casted as the extremum of an appropriate functional, and it is clear in this case that is it insensitive to field redefinitions (see e.g. \cite{Sen:2007qy} for a review). }  This is, in fact, already evident in our example. The two terms ${\cal L}_{\rm GB}$ and ${\cal L}_W$ are related by a field redefinition of the form (\ref{eq:redef}) and as we have observed the entropy of the Kerr black hole is the same for the two theories. 
In general, any  two Lagrangians related via \eqref{eq:redef} will give rise to the same  Wald entropy. 
We also note that $f(R)$ gravity theories are not related via field redefinitions to the action \eqref{eq:ba} with either ${\cal L}_{\rm GB}$ or ${\cal L}_W$, because the coefficient of $(R_{\mu\nu\alpha\beta})^2$ term is not modified by \eqref{eq:redef}. This is as well reflected in the fact that for these theories we obtained different results regarding both the Smarr relations and entropy products.

\subsection{A Lovelock Example}\label{sec:love}

Lovelock gravities \cite{Lovelock:1971yv} are distinguished among higher curvature theories by having equations of motion that depend only on the curvature tensor, and not on its derivatives; hence the equations are still only second order in derivatives of the metric.   
Above, we explored the effect of adding the quadratic Lovelock interaction ${\cal L}_{\rm GB}$ to the gravitational Lagrangian in $d=4$, finding that the entropy product picks up a dependence on the mass.  However, this case was in some sense too simple, with the new interaction term contributing to the black hole entropy formula (\ref{eq:w4d}), but not the dynamics of the theory.  The black hole solutions were still those of Einstein gravity.  Perhaps, once Lovelock interactions contribute to the dynamics and affect the black hole geometry, the mass independence of the entropy product relation might be restored.

The simplest way to assess this is to study the charged black holes of Einstein-Maxwell-Gauss-Bonnet theory in dimensions $d>4$.  In this case, fully explicit charged black hole solutions are known.  Charged black holes in general Lovelock gravity theories coupled to the Maxwell Lagrangian will be discussed in Appendix \ref{app:A}.
Here we take the action to be
\be\label{eq:da}
S={1\over 16\pi G} \int d^dx \sqrt{-g}\, \left({\cal L}_0+{\cal L}_1+{\cal L}_{\rm GB} \right) +   \int d^dx \sqrt{-g}\, {\cal L}_{\rm matter}~,
\ee
where as above ${\cal L}_1=R$ and ${\cal L}_{\rm matter}=(1/2)F_{\mu\nu}F^{\mu\nu}$, and we have also included a cosmological constant term ${\cal L}_0 =-2\Lambda$.
Taking the cosmological constant to be negative, it will also be convenient to define the equivalent AdS curvature radius $l$ and rescaled Gauss-Bonnet coupling $\alpha_d$ by
\be
\Lambda= -{(d-1)(d-2)\over 2\ell^2}~,\quad \alpha_d =(d-3)(d-4)\alpha~.
\ee

The Gauss-Bonnet term contributes to the equations of motion for spacetime dimensions $d>4$.
 Static black hole solutions to (\ref{eq:da}) have been known for some time, beginning with \cite{Boulware:1985wk,Wheeler:1985nh}.  Here for the most part we will follow conventions established  for Lovelock black holes used in \cite{Charmousis:2008kc} (see also  \cite{Garraffo:2008hu}). The static, charged black hole solutions to \eqref{eq:da} then have the form
\be\label{eq:metricGB}
ds^2=-V(r)dt^2+ {dr^2\over V(r)}+r^2 h_{ij} dx^i dx^j~,\quad F={q\over 4\pi r^{(d-2)}}dt\wedge dr~,
\ee
with the metric function
\be\label{eq:Vr}
V(r)=\kappa+ {r^2\over 2\alpha_d}\left[1+\epsilon \sqrt{1+4\alpha_d\left(-{1\over \ell^2}+{2\mu\over r^{d-1}}-{q^2\over r^{2(d-2)}}\right)}\right]~,
\ee
and $h_{ij}$  is a maximally symmetric space in $(d-2)$-dimensions which can be parametrized as
\be\label{eq:zz}
 h_{ij} dx^i dx^j={d\chi^2\over 1-\kappa \chi^2}+\chi^2 d\Omega_{d-3}^2 ~,
 \ee
 with $\kappa=0,\pm 1$ corresponding to a spherical,  flat or hyperbolic  horizons respectively. 
 The first notable difference with Einstein gravity is the appearance here of the parameter $\epsilon=\pm1$ in \eqref{eq:Vr}. 
This reflects that in the absence of a black hole ({\it i.e.} with $\mu=q=0$) and even in the limit $\Lambda=0$, the theory can have two different constant curvature vacuum solutions. 
Taking $\Lambda=0$, for $\epsilon=1$ the vacuum is  either dS or AdS space depending on the sign of $\alpha$, while for $\epsilon=-1$ we recover empty Minkowski space.  
The ADM mass $M$ and the electric charge $Q$ of the solution are given by
\be\label{eq:alphaQM}
 \mu={8\pi G M\over (d-2)\Sigma_{d-2}}~,\quad  q=\left({8\pi G\over \Sigma_{d-2} }\right)^{(d-3)/(d-2)} Q~, 
 \ee
 where $\Sigma_{d-2}$ is the area of $h_{ij}$ in \eqref{eq:zz}. 

It is sufficient for our purposes to restrict our focus further at this point to 
the case of asymptotically flat solutions in $d=5$ and with $\Lambda=0$ (see appendix \ref{app:A} for the more general case). 
The charged solutions  \eqref{eq:Vr} in this case reduces to
\be\label{eq:Vra}
V(r)=1+ {r^2\over 2\alpha_d}\left[1- \sqrt{1+4\alpha_d\left({2\mu\over r^{4}}-{q^2\over r^{6}}\right)}\right]~,
\ee
where we have set $\epsilon=-1$ and $\ell\to \infty$ and for simplicity considered only spherical horizons ($\kappa=1$).  These solutions  were  first found in \cite{Wiltshire:1985us,Wiltshire:1988uq}.
The roots of $V(r)$ are located at
\be
r_{\pm}^2 = {1\over 2}(2\mu-\alpha_d)\pm \sqrt{{1\over 4}(2\mu-\alpha_d)^2-q^2}~,
\ee
with $r_+$ denoting the outer horizon and $r_-$ the inner horizon.  Some useful relations are $(r_+r_-)^2=q^2$ and $r_+^2+r_-^2=2\mu-\alpha_d$.
The temperatures, electric potentials and entropies for each horizon are given by
\be\label{eq:T5D}
T_{\pm}
={1\over 2\pi}{r_+^2-r_-^2\over r_{\pm}(2\alpha_d +r_\pm^2)}~,\qquad
\Phi_{\pm}=3\left({\pi\over 4G}\right)^{1/3}{Q\over r_{\pm}^2} ~,\qquad S_{\pm}={\pi^2\over 2G}r_{\pm}^3\left(1+{6\alpha_d\over r_\pm^2}\right)~,
\ee
where the electric charge is given by $Q=\left({\pi\over 4G}\right)^{2/3}q$.

The product of the horizon entropies is then found to be
\be\label{eq:SS5D}
{S_{+}S_-\over 4\pi^2}= (1+{12\alpha_d \mu\over q^2} + { 30\alpha_d^2\over q^2})\, Q^3~,
\ee
which clearly depends on the mass through the parameter $\mu$.
It is worth noting that, if instead we evaluate the product of the horizon areas, then the result is unaffected by the presence of the Gauss-Bonnet term in the action.
The product of horizon areas $A_{\pm}=2\pi^2 r_{\pm}^3$ is given by
\be
{A_{+}A_-\over (8\pi)^2}= Q^3~.
\ee
which is the same result found in $d=5$ Einstein gravity.
Hence, while the product of horizon entropies is mass dependent, the product of horizon areas is actually quantized in $d=5$ Einstein-Maxwell-Gauss-Bonnet gravity as it was in $d=4$.  This is a curious result.  In view of the comments regarding field redefinitions in section (\ref{sec:fr}),  it is not clear whether this has physical significance, or should be regarded as coincidental. 

Finally, we record the first laws and Smarr formulas for both the inner and outer horizon quantities that hold for  these Einstein-Maxwell-Gauss-Bonnet
black holes.   For the extended first laws which allow for the variation of the Gauss-Bonnet,  we obtain
\be
dM= \pm T_\pm dS_\pm + \Phi_\pm dQ+ 
\Theta^{(2)}_\pm {d\alpha_d}  ~,
\ee
while the Smarr relations are given by
\be
 M=\pm {3\over 2}  T_\pm S_\pm + \Phi_\pm Q+\Theta^{(2)}_\pm  {\alpha_d}  ~,
\ee
and one finds that  $\Theta^{(2)}_\pm ={3\pi \over 8G} (1\mp 8\pi \alpha_d r_\pm T_\pm)$
is the potential conjugate to the Gauss-Bonnet coupling $\alpha_d$. As in the $d=4$ case above, it is necessary to include the  $d\alpha_d$ term in the first law in order to obtain the correct Smarr formula via an overall rescaling \cite{Kastor:2010gq}.  We note that relations of the form \eqref{manyrelations} between the potentials at the inner and outer horizons can also be obtained in this case.

\section{Discussion}\label{sec:disc}

We have found that the mass independence of the horizon entropy product  is sensitive to exactly which higher curvature terms appear in the Lagrangian.  Simple counter-examples have shown that the mass independent form given in  equation \eqref{eq:main} is in fact {\it not} universal. Given that there exists a broad and physically important class of theories (and solutions) for which \eqref{eq:main} holds, it becomes a puzzle as to what specific features of the gravitational theory are required for this to be the case. 

One possible piece of this puzzle is the fact that 
the form of the Smarr relation is modified for all the ``failures'' we have studied, while it remains unchanged for the ``successes''. There is no obvious relation between the  horizon entropy products and the Smarr relations.  Nevertheless, in the ``failing" theories both are drastically modified by the presence of higher curvature terms.  Another possible puzzle piece is the continued mass independence of the horizon area product in all our higher curvature examples, even though its thermodynamic interpretation becomes obscure.  Alternatively, it may be that the relations \eqref{manyrelations} between  thermodynamic quantities at the inner and outer horizons, whose existence does not depend on the mass independence of the horizon entropy product, may prove to be of fundamental significance.

The fact that the horizon entropy product is not generically independent of the mass has important implications.  One immediate  consequence is the breakdown of a CFT description of the black hole proposed in \cite{CMS} when higher curvature terms are present. The success of this program relies heavily on the mass independence of the horizon entropy product as noted in \cite{Chen:2012mh,Chen:2013rb}.  In particular, equation  \eqref{eq:main} insures that the central charge of the 2D CFT is similarly independent of the mass, and in turn that it will agree with the value obtained using Kerr/CFT techniques (under certain assumptions). For examples in which the mass independence of the product $S_+S_-$ is no longer protected in the presence of higher curvature terms,  the conclusion would be that only the low energy description of the thermodynamics is captured by a Cardy formula. 

It would clearly be of interest  to test mass independence of the product $S_+S_-$ in the context of string theory black holes. The class of higher derivative corrections that appear in this context are significantly more complicated than those considered here.  However, it is possible to construct BPS  solutions in these theories and compute the corrections to the Wald entropy (see e.g. \cite{Mohaupt:2000mj, Sen:2007qy,Castro:2008ne} and references therein). The advantage in this case is that exists a better understanding of the microscopic degrees of freedom for BPS configurations than is available for the simpler higher curvature theories considered here.   It would be interesting to investigate the properties of finite temperature black holes in these theories and understand what role the entropy product  $S_+S_-$ plays in the D-brane construction.

Finally, a geometric derivation of \eqref{eq:main} should continue to be a goal.   In addition to the implications for  the statistical interpretation of the Wald entropy in terms of a two dimensional  CFT, failure of the mass independence of the horizon entropy product may reflect some other potential pathologies of a given classical higher curvature theory. It would be interesting if mass independence of the horizon entropy product  could serve as a simple diagnostic test for consistency of higher curvature terms,  associating $S_+S_-\neq S_{\rm ext}^2$ with a potential sickness of some sort.   For example, It might possibly reflect thermodynamic instability or perhaps capture the potential for classical hair. The verdict remains unclear.

\section*{Acknowledgements}

{We thank St\'{e}phane Detournay, Ted Jacobson, Hideki Maeda, Sameer Murthy, Gim Seng Ng, Julio Oliva, Loganayagam Ramalingam,  Mar\'{\i}a J. Rodr\'{\i}guez, Jennie Traschen and Stefan Vandoren for useful conversations. AC's work is supported by the Fundamental Laws Initiative of the Center for the Fundamental Laws of Nature, Harvard University. GG's work is supported by grants of ANPCyT, CONICET, and UBA.}

\appendix

\section{Conventions and Definitions}\label{sec:defn}

In this section we will collect some definitions that we used throughout our derivations. 
For most computations it is convenient to use the ADM  form of the metric, where  
\bea
ds^2= -N^2dt^2 + \gamma_{ab}(dx^a + N^a dt )(dx^b + N^b dt )~,
\eea
with $x^a$ spatial directions and $a,b=1,\ldots,d-1$, and the greek indices span all spacetime directions $\mu,\nu=0,1,\ldots, d-1$. $N(x^a)$ and $N^b(x^a)$ are the lapse function and the shift vector respectively. 

 The horizons are defined as the zeroes of the appropriate radial component of the metric.  We use the notation $r=r_i$ to refer to the location of these horizons; and we emphasize that $r_i$ is not restricted to be a real number. For those case where there are only two horizons we use the notation   $r=r_\pm$ and refer to them as  inner ($r_-$) and outer ($r_+$) horizon.

 As in \cite{Castro:2012av}, the angular potentials and temperatures for each horizon are 
\bea\label{genpot}
\Omega_{k}^{i} =- \left. N^{k}\right|_{r_i}   ~,\quad T_{i} ={1\over 4\pi} \left| {(N^2)' \over \sqrt{g_{rr}N^2}}\right|_{r_{i}}~,
\eea
where $k=1,2,...,[\frac{d-1}{2}]$. For a stationary solution, the null Killing vectors that define the inner and outer horizon are then
\bea
\chi_{i}=\partial_t -\Omega_{k}^{i}\,\partial_{\phi^k}~,
\eea
where the coordinates $\phi^k$ has periodicity $2\pi$.  And finally, the electric potential is defined as
\be
\label{elecpot}
\Phi_{i}= (\chi^\mu A_\mu)_\infty -  (\chi^\mu A_\mu)_{ r_{i}}~.
\ee
The normalization of $\Phi_{i}$ depends on our definition of electric charge; usually we will use units where the electric charge is quantized in Planck units when appropriate.

All conserved charges are computed using Komar integrals, and we follow the conventions in \cite{Wald:1993nt,Iyer:1994ys}. For example, for Einstein gravity the mass is given by 
\be
M=-{1\over 8\pi}\int_{\Sigma_{d-2}} d\Sigma^{\mu_3\cdots \mu_d}\epsilon_{\mu_1\mu_2 \mu_3 \cdots \mu_{d} } \nabla^{\mu_1} \xi^{\mu_2} ~.
\ee
where $\xi$ is a time-like Killing vector normalized such that at infinity $|\xi|^2\to-1$. And for the angular momentum
\be
J={1\over 16\pi}\int_{\Sigma_{d-2}}d\Sigma^{\mu_3\cdots \mu_d} \epsilon_{\mu_1\mu_2 \mu_3 \cdots \mu_{d} } \nabla^{\mu_1} \eta^{\mu_2}  ~.
\ee
where $\eta$ is the spatial Killing direction. 

The Wald entropy is given by 
\be\label{eq:wald}
S_\pm= 2\pi \int_{\Sigma_{d-2}^{i}} d\Sigma {\partial {\cal L}\over \partial R_{\mu\nu\lambda\rho}}\epsilon_{\mu\nu}\epsilon_{\lambda\rho} ~,
\ee
where $\Sigma_{d-2}^{i}$ is a spacelike bifurcation surface at the virtual horizon ($r_i$) and $\epsilon^{\mu\nu}$ is the binormal to this surface $\Sigma_{d-2}^{i}$.

\section{General Lovelock Black Holes}\label{app:A}

It is rather straight forward to extend our analysis in section \ref{sec:love} to general static solutions of Einstein-Maxwell-Lovelock theories with arbitrary Lovelock coefficients $\hat\alpha_i$ (including cosmological constant) and in arbitrary dimensions. 

The action will be the general Lovelock action plus the Maxwell action in $d$ dimensions \cite{Charmousis:2008kc}:
\be
S={1\over 16\pi G}\int \mathcal{L}+{1\over 4}\int F\wedge *F~,
\ee
where $ F=dA$ and the Lovelock terms are
\be
\mathcal{L}=\sum_{p=0}^{[d/2]} \hat\alpha_p \mathcal{L}_p~,\quad \mathcal{L}_p\equiv \epsilon_{a_1b_1\cdots c_j} \bigwedge_{i=0}^p R^{a_i b_i}\bigwedge_{j=0}^{d-2p} e^{c_j} ,  
\ee
In this notation, the term $p=0$ corresponds to the volume element which will give rise to a cosmological constant; hence $\hat\alpha_0 =  \Lambda$.  The term $p=1$ is simply the Einstein-Hilbert Lagrangian, where $\hat\alpha_1=1$. The last term in the sum, i.e. $p=[d/2]$, always corresponds to a topological invariant in $d$ dimensions and hence it does not affect the equations of motion.  If we take $d=4$, we are back again to Gauss-Bonet, the same theory as we studied in section \ref{sec:4d}. In $d=5$ we are just adding the term $p=2$, which reduces to our example in section \ref{sec:love} with $\hat \alpha_0=0$.
To keep track of the dependence on the cosmological constant we introduce the  notation
\be
\mathcal{L}=\sum_{p=s}^{[d/2]} \hat\alpha_p \mathcal{L}_p, \quad s=\left\{ \begin{array}{ll}
0& \mbox{if}~~ \Lambda\ne0~,\\
1& \mbox{if}~~ \Lambda=0~.
\end{array}\right.
\ee


Static solutions with both mass and electric charge are as follows. The metric and field strength are
\be 
ds^2= -V(r)dt^2 + {dr^2\over V(r)}+r^2  h_{ij} dx^i dx^j~,\quad F= {\hat q\over r^{d-2}} dt\wedge dr ~,
\ee
with   
\be\label{eq:VV}
V(r)= \kappa -r^2 f(r)~,
\ee
and $h_{ij}$ is given by \eqref{eq:zz} where the normalized horizon curvature takes values $\kappa=0,\pm 1$. 
The function $f(r)$ is a solution to the polynomial equation 
\be\label{eq:PP}
P(f)= \sum_{p=0}^{[(d-1)/2]} \alpha_p f^p={2\mu \over r^{d-1}}- {q^2\over r^{2(d-2)}}~.  
 \ee
 with 
\be
q^2={8\pi G\over (d-2)(d-3)} \hat q^2~,
\ee
and
 \be
 \alpha_0={\hat \alpha_0\over \hat \alpha_1} {1\over (d-1)(d-2)}~,\quad \alpha_1=1~,\quad \alpha_p= {\hat\alpha_p\over \hat \alpha_1}\prod_{n=3}^{2p}(d-n)~,\quad p\geq2~.
 \ee
 Here $\mu$  and  $q$ are proportional to the ADM mass and electric charge, with the normalizations given by \eqref{eq:alphaQM}.
For generic values of $\alpha_k$,  it is difficult to solve explicitly for $f(r)$. Still, the fact that the solution fits into this polynomial form simplifies many things. As we will see, we do not need to solve this equation for finding the entropy. 

The location of the horizons $r_i$ are determined by
\be 
V(r_i)=0=\kappa- r_i^2 f(r_i)~.
\ee
Plugging this into  (\ref{eq:PP}) we have
\be
P\left[\kappa r_i^{-2}\right]=2 \mu r_i^{-d+1} - q^2 r_i^{2(2-d)}~.
\ee
Multiplying by $r_i^{2(d-2)}$ gives
\begin{align}
\sum_{p=0}^{[(d-1)/2]} \alpha_p \kappa^p r_i^{2(d-p-2)}-2\mu r_i^{d-3}+q^2 =0~.\label{eq:HorPoly}
\end{align}
The degree of this polynomial depends on the highest power $2(d-p-2)$. If $\Lambda\ne0$ ( $\alpha_0\ne0$) the degree is $2(d-2)$; for $\Lambda=0$ ($\alpha_0=0$) the degree is $2(d-3)$. Since  equation \eqref{eq:HorPoly} is just a polynomial we may write it in its factorized form
\begin{equation}
\sum_{p=s}^{[(d-1)/2]} \alpha_p \kappa^p r^{2(d-p-2)}-2\mu r^{d-3}+q^2 = {\alpha_s }\kappa^s\prod_{i=1}^{2(d-s-2)}(r-r_i)~.\label{eq:roots}
\end{equation}
Generically, there will be $2(d-s-2)$ horizons. When $s=0$, $\Lambda\ne0$, many of the horizon radii will be complex and will correspond to virtual horizons;  however, we are assuming that at least one horizon is a real positive number. If $\kappa=0$, assuming that $\alpha_0\neq 0$,  the polynomial  remains of the same  degree. If $\alpha_0=\kappa=0$  the situation is degenerate and the analysis below does not apply. 

Comparing the left and right hand side of \eqref{eq:roots} gives some useful relations between the global charges $(\mu,q)$ and locations of the horizon. In particular we find
\be\label{eq:aaDD}
{q^2\over \alpha_s \kappa^s} = \prod_{k=i}^{2(d-s-2)} r_{i} ~.
\ee
This will imply that the product of the areas is always independent of the mass $\mu$.

Each root $r_i$ will satisfy it's own ``thermodynamical'' relation. Some of these thermodynamic quantities are straightforward to compute. For instance, from  \eqref{genpot} the temperature is given by
\bea
T_{i}&=&{V'(r_i)\over 4\pi}\cr
&=& {1\over 4\pi r_i P\rq{}(\kappa r^{-2}_i)}\left[\kappa \sum_p (d-2p-1) \alpha_p(\kappa r_i^{-2})^{p-1}+ (3-d)q^2 r_i^{2(3-d)} \right] 
\eea
where we used $P\rq{}(x)= \sum_p p\alpha_p x^{p-1}$ and $2\mu= r^{d-1}_i P(\kappa r^{-2}_i) +q^2 r_i^{3-d}$.
For $d=5$ we have  $\hat \alpha_0=0$, $\hat \alpha_1=1$, $\hat \alpha_2 =\alpha$, and this yields
\be
T_i= {r_i^4- q^2  \over 2\pi r_i^3 (2 \alpha_d +r_i^2)}~,
\ee
which is in exact agreement with our calculation in (\ref{eq:T5D}). 

The entropy associated to each horizon is given by
\bea\label{eq:SLV}
S_i&=& {\Sigma_{d-2} \over 4G} \sum_{p=s}^{[d/2]} {p(d-2)\over d-2p} \alpha_p \kappa^{p-1} r_i^{d-2p}\cr 
&=&{A_i \over 4G}\, \sum_{p=s}^{[d/2]} {p(d-2)\over d-2p} \alpha_p \left({\kappa\over r_i^2}\right)^{p-1} ~,
\eea  
where $\Sigma_{d-2}$ is the area of $h_{ij}$, and $A_i=\Sigma_{d-2}r_i^{d-2}$ is the area of each horizon. This result can be obtained either by using Wald's formula \eqref{eq:wald} or the thermodynamic method in  \cite{Jacobson:1993xs}.\footnote{We are including in \eqref{eq:SLV} the contribution of the topological term given by $p=[d/2]$. } For the Gauss-Bonnet theory ($\alpha_k=0$ for $k>2$), \eqref{eq:SLV} reduces to 
\be\label{eq:SBGx} 
S_i={A_i\over 4G}\left[\alpha_1+2\alpha_2 {(d-2)\over (d-4)}\kappa r_i^{-2}\right] ~.\ee
which reduces to \eqref{eq:SS5D}.

To find the product of the entropies (\ref{eq:SLV}) for all horizons it is convenient to write the entropy as 
\begin{align}
S_i&= {A_i \over 4 G r_i^{2([d/2]-1)}} \prod_{n=1}^{[d/2]-1} \pr{r_i^2-C_n}~,
\end{align}
for appropriate choices of the coefficients $C_n$ such that we recover \eqref{eq:SLV}. Notice that $C_n$ only depends on $\alpha_p$, $\kappa$ and $d$, and it does not depend on either $\mu$ or $q$. 
Using \eqref{eq:roots} with $r\to \pm C_n$ we can write
\begin{equation}
\prod_{i=1}^{2(d-s-2)}(r_i\pm \sqrt{C_n})= {1\over \alpha_s \kappa^s}\left[\sum_{p=s}^{[(d-1)/2]} \alpha_p \kappa^p C_n^{d-p-2}-\mu \pr{\pm  \sqrt{C_n}}^{d-3}+q^2 \right]~. \label{eq:prodr-ia}
\end{equation}
And the product of them for all roots $r_i$ is
\begin{align}
 \mathcal{G}_d\equiv &\prod_{n=1}^{[d/2]-1}\prod_{k=1}^{2(d-s-2)}(r_k-\sqrt{C_n})\prod_{j=1}^{2(d-s-2)}(r_j+\sqrt{C_n})=\cr
 \mbox{Even\ }d:\quad &\prod_{n=1}^{[d/2]-1}   {1\over \alpha_s^2 \kappa^{2s}}\left[\left\{\sum_{p=s}^{[(d-1)/2]} \alpha_p \kappa^p C_n^{d-p-2}+q^2\right\}^2 -\mu^2 C_n^{d-3}\right] \cr
  \mbox{Odd\ }d:\quad &\prod_{n=1}^{[d/2]-1}   {1\over \alpha_s^2 \kappa^{2s}}\left[\sum_{p=s}^{[(d-1)/2]} \alpha_p \kappa^p C_n^{d-p-2}+q^2 -\mu C_n^{(d-3)/2}\right]^2 \label{eq:GGen}
\end{align}
Therefore  the entropy product above becomes:
\begin{align}\label{eq:prodS}
\prod_{i=1}^{2(d-s-2)}  S_i&=\pr{{\Sigma_{d-2} \over 4G}}^{2(d-s-2)} \pr{{q^2\over \alpha_s \kappa^s}}^{(d-2[d/2])} \mathcal{G}_d 
\end{align}
Thus the product of the entropies will have a mass dependent term.

Looking at $\mathcal{G}_d$ in  (\ref{eq:GGen})  we see that it will contain a $\mu$ dependent term, unless all $C_i$ vanish. From (\ref{eq:SLV}) we see that this happens for a generic Lovelock theory if $\kappa$, the normalized horizon curvature, vanishes. This as well implies that  for planar black holes there are no correction to Bekenstein-Hawking area law, as noted in \cite{Dehghani:2002wn}.  The product of the entropy (per unit area)  satisfies
\be
\prod_i {S_i}= \prod_i {A_i\over 4G \Sigma_{d-2}} = \left({q^2\over16 G^2 \alpha_0}\right)^{d-2}~.
\ee

\section{Yang Monopole Solution}\label{app:YM}

In this appendix we  discuss the self-gravitating Yang monopole in $D=6$ dimensions \cite{Gibbons:2006wd}. This
is a solution of Einstein gravity coupled to Yang-Mills theory for the group $%
SU(2)$. The action of the theory is%
\[
S=\frac{1}{16\pi G}\int d^{6}x\sqrt{-g}(\mathcal{L}_{0}+\mathcal{L}_{1})-%
\frac{1}{2g^{2}}\int d^{6}x\sqrt{-g}\text{tr}(F^{2})~,
\]%
with the gauge field $A$%
\[
A=A_{\mu }^{a}T_{a}dx^{\mu }~,
\]%
with $T_{a}$ being the generators of the Lie algebra. 

The one-form of the Yang monopole configuration is%
\[
A=\frac{\Sigma _{ij}n^{i}n^{j}}{1+\sqrt{1-n^{2}}}~,
\]%
where $n^{i}$ are coordinates on the four sphere ($i=1,2,3,4$) such that the
metric on the unit four-sphere reads%
\[
d\Omega _{4}^{2}=\left( \delta _{ij}+\frac{n_{i}n_{j}}{1-n_{k}n^{k}}\right)
dx^{i}dx^{j}~,
\]
with $n_{i}=\delta _{ij}n^{j}$. The constants $\Sigma _{ij}=\eta _{ij}^{a}T_{a}$ obey 
\[
\eta _{ij}^{a}\eta _{kl}^{b}f_{ab}^{c}=\delta _{li}\eta _{jk}^{c}+\delta
_{kj}\eta _{il}^{c}-\delta _{ki}\eta _{jl}^{c}-\delta _{lj}\eta _{ik}^{c}~,
\]
where $f_{ab}^c$ are the structure constant and satisfy $\eta _{\lbrack ij}^{a}\eta _{kl]}^{a}=\varepsilon _{ijkl}$.  The form of the metric charged under $A$ takes the form
\[
ds^{2}=-V(r)dt^{2}+\frac{dr^{2}}{V(r)}+r^{2}d\Omega _{4}^{2}~,
\]%
with%
\be\label{eq:vrym}
V(r)=1-\frac{M}{r^{3}}-\frac{Q^{2}}{r^{2}}-\frac{\Lambda }{10}r^{2},
\ee%
for $M$ and $Q$ real constants. Notice that this solution exhibits a charge
term $-Q^{2}/r^{2}$ that damps off slower than the six-dimensional Newtonian
mass term $-M/r^{3}$. This is related to the fact that self-gravitating Yang
monopole has infinite energy \cite{Gibbons:2006wd}. 

This will be another example which shows peculiar behavior regarding the product of the entropies: For each horizon we assign entropy $S_i=A_i/4G$. The product of the entropies for $\Lambda\neq0$ is
\[\prod_{i=1}^5 S_i=\prod_{i=1}^5 {A_i\over 4G}=\pr{{\Sigma_4\over 4G}}^5\pr{{ 10 M\over \Lambda}}^4~, \]
which is independent of the charge $Q$ while does depend on $M$. 

Similarly, when $\Lambda =0$ the radii of the horizons are located at zeroes
of $V(r)$. There are three zeroes and the product of the entropies becomes
\begin{equation}
\prod_{i=1}^3 S_i=\pr{{\Sigma_4\over 4G}}^3 M^4= \prod_{i=1}^3 S_{i,\rm ext}~, \label{producto}
\end{equation}
 where $S_{i,\rm ext}$ is the entropy of each virtual horizon at zero temperature (i.e when two roots of \eqref{eq:vrym} coincide). It is rather peculiar that the product is independent of the charge $Q$. However, we emphasize that the direct identification of $M$ and $Q$ with the mass and the electric charge is misleading. In fact, the solution presents infinite mass if $Q\neq 0$; that is, $M$ corresponds to the ADM case if $Q=0$ where the product of entropies (\ref{producto}) vanishes. It is also very interesting that the product of the entropies equals the extremal ($T=0$) entropy.

\nopagebreak


\begin{thebibliography}{42} 
  
\bibitem{Wald:1993nt} 
  R.~M.~Wald,
  ``Black hole entropy is the Noether charge,''
  Phys.\ Rev.\ D {\bf 48}, 3427 (1993)
  [gr-qc/9307038].
  

\bibitem{Iyer:1994ys} 
  V.~Iyer and R.~M.~Wald,
  ``Some properties of Noether charge and a proposal for dynamical black hole entropy,''
  Phys.\ Rev.\ D {\bf 50}, 846 (1994)
  [gr-qc/9403028].
  
\bibitem{Sen:2007qy} 
  A.~Sen,
  ``Black Hole Entropy Function, Attractors and Precision Counting of Microstates,''
  Gen.\ Rel.\ Grav.\  {\bf 40}, 2249 (2008)
  [arXiv:0708.1270 [hep-th]].


\bibitem{Cvetic:2010mn}
  M.~Cvetic, G.~W.~Gibbons, C.~N.~Pope,
  ``Universal Area Product Formulae for Rotating and Charged Black Holes in Four and Higher Dimensions,''
  Phys.\ Rev.\ Lett.\  {\bf 106 } (2011)  121301.
  [arXiv:1011.0008 [hep-th]].

  
\bibitem{Castro:2012av} 
  A.~Castro and M.~J.~Rodriguez,
  ``Universal properties and the first law of black hole inner mechanics,''
  Phys.\ Rev.\ D {\bf 86}, 024008 (2012)
  [arXiv:1204.1284 [hep-th]].

\bibitem{Detournay:2012ug}
  S.~Detournay,
  ``Inner Mechanics of 3d Black Holes,''
  Phys.\ Rev.\ Lett.\  {\bf 109} (2012) 031101
  [arXiv:1204.6088 [hep-th]].
  
  
\bibitem{Chen:2012mh}
  B.~Chen, S.~-x.~Liu, and J.~-j.~Zhang,
  ``Thermodynamics of Black Hole Horizons and Kerr/CFT Correspondence,''
  JHEP {\bf 1211} (2012) 017
  [arXiv:1206.2015 [hep-th]].

\bibitem{Chen:2013rb} 
  B.~Chen, Z.~Xue and J.~-j.~Zhang,
  ``Note on Thermodynamic Method of Black Hole/CFT Correspondence,''
  arXiv:1301.0429 [hep-th].

\bibitem{Castro:2013kea} 
  A.~Castro, J.~M.~Lapan, A.~Maloney and M.~J.~Rodriguez,
  ``Black Hole Monodromy and Conformal Field Theory,''
  arXiv:1303.0759 [hep-th].

\bibitem{Strominger:1997eq} 
  A.~Strominger,
  ``Black hole entropy from near horizon microstates,''
  JHEP {\bf 9802}, 009 (1998)
  [hep-th/9712251].
  
\bibitem{Larsen:1997ge}
 F.~Larsen,
  ``A String model of black hole microstates,''
 Phys.\ Rev.\  {\bf D56 } (1997)  1005-1008.
 [hep-th/9702153].


\bibitem{Guica:2008mu} 
  M.~Guica, T.~Hartman, W.~Song and A.~Strominger,
  ``The Kerr/CFT Correspondence,''
  Phys.\ Rev.\ D {\bf 80}, 124008 (2009)
  [arXiv:0809.4266 [hep-th]].


\bibitem{CMS}
A. Castro, A. Maloney and A. Strominger,
`` Hidden conformal symmetry of the Kerr black hole,''
Phys. Rev. {\bf D82}, 024008 (2010), arXiv:1004.0996 [hep-th].
  
\bibitem{Lovelock:1971yv} 
  D.~Lovelock,
  ``The Einstein tensor and its generalizations,''
  J.\ Math.\ Phys.\  {\bf 12}, 498 (1971).
  
\bibitem{Curir1} 
A.~Curir,
``Spin entropy of a rotating black hole"
Nuovo Cimento B,{\bf 51B},  262 (1979).

\bibitem{Curir2} 
A.~Curir and M.~ Francaviglia,
``Spin thermodynamics of a Kerr black hole"
Nuovo Cimento B,{\bf 52B},  165 (1979).

 

\bibitem{Ansorg:2008bv} 
  M.~Ansorg and J.~Hennig,
  ``The Inner Cauchy horizon of axisymmetric and stationary black holes with surrounding matter,''
  Class.\ Quant.\ Grav.\  {\bf 25}, 222001 (2008)
  [arXiv:0810.3998 [gr-qc]].
  
\bibitem{Ansorg:2009yi} 
  M.~Ansorg and J.~Hennig,
  ``The Inner Cauchy horizon of axisymmetric and stationary black holes with surrounding matter in Einstein-Maxwell theory,''
  Phys.\ Rev.\ Lett.\  {\bf 102}, 221102 (2009)
  [arXiv:0903.5405 [gr-qc]].

\bibitem{Meessen:2012su} 
  P.~Meessen, T.~Ortin, J.~Perz and C.~S.~Shahbazi,
  ``Black holes and black strings of N=2, d=5 supergravity in the H-FGK formalism,''
  JHEP {\bf 1209}, 001 (2012)
  [arXiv:1204.0507 [hep-th]].

\bibitem{Visser:2012zi}
  M.~Visser,
  ``Quantization of area for event and Cauchy horizons of the Kerr-Newman black hole,''
  JHEP {\bf 1206} (2012) 023
  [arXiv:1204.3138 [gr-qc]].


  
  
\bibitem{Faraoni:2012je}
  V.~Faraoni and A.~F.~Z.~Moreno,
  ``Are quantization rules for horizon areas universal?,''
  arXiv:1208.3814 [hep-th].

\bibitem{Toldo:2012ec} 
  C.~Toldo and S.~Vandoren,
  ``Static nonextremal AdS4 black hole solutions,''
  JHEP {\bf 1209}, 048 (2012)
  [arXiv:1207.3014 [hep-th]].

\bibitem{Klemm:2012vm}
  D.~Klemm and O.~Vaughan,
  ``Nonextremal black holes in gauged supergravity and the real formulation of special geometry II,''
  Class.\ Quant.\ Grav.\  {\bf 30} (2013) 065003
  [arXiv:1211.1618 [hep-th]].


\bibitem{Visser:2012wu}
  M.~Visser,
  ``Area products for black hole horizons,''
  arXiv:1205.6814 [hep-th].

\bibitem{Gnecchi:2012kb}
  A.~Gnecchi and C.~Toldo,
  ``On the non-BPS first order flow in N=2 U(1)-gauged Supergravity,''
  JHEP {\bf 1303} (2013) 088
  [arXiv:1211.1966 [hep-th]].

  
\bibitem{Banados:1992wn} 
  M.~Banados, C.~Teitelboim and J.~Zanelli,
  ``The Black hole in three-dimensional space-time,''
  Phys.\ Rev.\ Lett.\  {\bf 69}, 1849 (1992)
  [hep-th/9204099].
  
\bibitem{Banados:1992gq} 
  M.~Banados, M.~Henneaux, C.~Teitelboim and J.~Zanelli,
  ``Geometry of the (2+1) black hole,''
  Phys.\ Rev.\ D {\bf 48}, 1506 (1993)
  [gr-qc/9302012].
  
\bibitem{Brown:1986nw} 
  J.~D.~Brown and M.~Henneaux,
  ``Central Charges in the Canonical Realization of Asymptotic Symmetries: An Example from Three-Dimensional Gravity,''
  Commun.\ Math.\ Phys.\  {\bf 104}, 207 (1986).
  
\bibitem{Saida:1999ec} 
  H.~Saida and J.~Soda,
  ``Statistical entropy of BTZ black hole in higher curvature gravity,''
  Phys.\ Lett.\ B {\bf 471}, 358 (2000)
  [gr-qc/9909061].


\bibitem{Kraus:2005vz} 
  P.~Kraus and F.~Larsen,
  ``Microscopic black hole entropy in theories with higher derivatives,''
  JHEP {\bf 0509}, 034 (2005)
  [hep-th/0506176].

\bibitem{Gupta:2007th} 
  R.~K.~Gupta and A.~Sen,
  ``Consistent Truncation to Three Dimensional (Super-)gravity,''
  JHEP {\bf 0803}, 015 (2008)
  [arXiv:0710.4177 [hep-th]].
  
\bibitem{Bergshoeff:2009hq} 
  E.~A.~Bergshoeff, O.~Hohm and P.~K.~Townsend,
  ``Massive Gravity in Three Dimensions,''
  Phys.\ Rev.\ Lett.\  {\bf 102}, 201301 (2009)
  [arXiv:0901.1766 [hep-th]].

\bibitem{Clement:2009gq} 
  G.~Clement,
  ``Warped AdS(3) black holes in new massive gravity,''
  Class.\ Quant.\ Grav.\  {\bf 26}, 105015 (2009)
  [arXiv:0902.4634 [hep-th]].

\bibitem{Sheykhi:2012zz} 
  A.~Sheykhi,
  ``Higher-dimensional charged $f(R)$ black holes,''
  Phys.\ Rev.\ D {\bf 86}, 024013 (2012)
  [arXiv:1209.2960 [hep-th]].
  
\bibitem{Sotiriou:2008rp} 
  T.~P.~Sotiriou, V.~Faraoni and ,
  ``f(R) Theories Of Gravity,''
  Rev.\ Mod.\ Phys.\  {\bf 82}, 451 (2010)
  [arXiv:0805.1726 [gr-qc]].

\bibitem{Moon:2011hq} 
  T.~Moon, Y.~S.~Myung and E.~J.~Son,
  ``f(R) black holes,''
  Gen.\ Rel.\ Grav.\  {\bf 43}, 3079 (2011)
  [arXiv:1101.1153 [gr-qc]].
  
\bibitem{Kastor:2009wy} 
  D.~Kastor, S.~Ray and J.~Traschen,
  ``Enthalpy and the Mechanics of AdS Black Holes,''
  Class.\ Quant.\ Grav.\  {\bf 26}, 195011 (2009)
  [arXiv:0904.2765 [hep-th]].

\bibitem{Kastor:2010gq} 
  D.~Kastor, S.~Ray and J.~Traschen,
  ``Smarr Formula and an Extended First Law for Lovelock Gravity,''
  Class.\ Quant.\ Grav.\  {\bf 27}, 235014 (2010)
  [arXiv:1005.5053 [hep-th]].
  
\bibitem{Cvetic:2010jb} 
  M.~Cvetic, G.~W.~Gibbons, D.~Kubiznak and C.~N.~Pope,
  ``Black Hole Enthalpy and an Entropy Inequality for the Thermodynamic Volume,''
  Phys.\ Rev.\ D {\bf 84}, 024037 (2011)
  [arXiv:1012.2888 [hep-th]].
  
\bibitem{Dolan:2013ft} 
  B.~P.~Dolan, D.~Kastor, D.~Kubiznak, R.~B.~Mann and J.~Traschen,
  ``Thermodynamic Volumes and Isoperimetric Inequalities for de Sitter Black Holes,''
  arXiv:1301.5926 [hep-th].
  
\bibitem{Kastor:2011qp} 
  D.~Kastor, S.~Ray, J.~Traschen and ,
  ``Mass and Free Energy of Lovelock Black Holes,''
  Class.\ Quant.\ Grav.\  {\bf 28}, 195022 (2011)
  [arXiv:1106.2764 [hep-th]].




\bibitem{Okamoto}
  I.~Okamoto,  and O.~Kaburaki, ``The 'inner-horizon thermodynamics' of Kerr black holes,'' Monthly Notices of the Royal Astronomical Society {\bf 255}, 539 (1992).

\bibitem{Jacobson:1993vj} 
  T.~Jacobson, G.~Kang and R.~C.~Myers,
  ``On black hole entropy,''
  Phys.\ Rev.\ D {\bf 49}, 6587 (1994)
  [gr-qc/9312023].

  
\bibitem{Boulware:1985wk} 
  D.~G.~Boulware and S.~Deser,
  ``String Generated Gravity Models,''
  Phys.\ Rev.\ Lett.\  {\bf 55}, 2656 (1985).
  
\bibitem{Wheeler:1985nh} 
  J.~T.~Wheeler,
  ``Symmetric Solutions to the Gauss-Bonnet Extended Einstein Equations,''
  Nucl.\ Phys.\ B {\bf 268}, 737 (1986).

\bibitem{Charmousis:2008kc} 
  C.~Charmousis,
  ``Higher order gravity theories and their black hole solutions,''
  Lect.\ Notes Phys.\  {\bf 769}, 299 (2009)
  [arXiv:0805.0568 [gr-qc]].
  
\bibitem{Garraffo:2008hu} 
  C.~Garraffo and G.~Giribet,
  ``The Lovelock Black Holes,''
  Mod.\ Phys.\ Lett.\ A {\bf 23}, 1801 (2008)
  [arXiv:0805.3575 [gr-qc]].




\bibitem{Wiltshire:1985us} 
  D.~L.~Wiltshire,
  ``Spherically Symmetric Solutions Of Einstein-maxwell Theory With A Gauss-Bonnet Term,''
  Phys.\ Lett.\ B {\bf 169}, 36 (1986).
  
\bibitem{Wiltshire:1988uq} 
  D.~L.~Wiltshire,
  ``Black Holes In String Generated Gravity Models,''
  Phys.\ Rev.\ D {\bf 38}, 2445 (1988).

\bibitem{Mohaupt:2000mj} 
  T.~Mohaupt,
  ``Black hole entropy, special geometry and strings,''
  Fortsch.\ Phys.\  {\bf 49}, 3 (2001)
  [hep-th/0007195].

\bibitem{Castro:2008ne} 
  A.~Castro, J.~L.~Davis, P.~Kraus, F.~Larsen and ,
  ``String Theory Effects on Five-Dimensional Black Hole Physics,''
  Int.\ J.\ Mod.\ Phys.\ A {\bf 23}, 613 (2008)
  [arXiv:0801.1863 [hep-th]].
  
\bibitem{Jacobson:1993xs} 
  T.~Jacobson and R.~C.~Myers,
  ``Black hole entropy and higher curvature interactions,''
  Phys.\ Rev.\ Lett.\  {\bf 70}, 3684 (1993)
  [hep-th/9305016].


\bibitem{Dehghani:2002wn} 
  M.~H.~Dehghani,
  ``Charged rotating black branes in anti-de Sitter Einstein-Gauss-Bonnet gravity,''
  Phys.\ Rev.\ D {\bf 67}, 064017 (2003)
  [hep-th/0211191].

  
\bibitem{Gibbons:2006wd} 
  G.~W.~Gibbons and P.~K.~Townsend,
  ``Self-gravitating Yang Monopoles in all Dimensions,''
  Class.\ Quant.\ Grav.\  {\bf 23}, 4873 (2006)
  [hep-th/0604024].
  
 

  
 
\end{thebibliography}
\end{document}